\documentclass[reprint,showpacs,showkeys,amsmath,amssymb,floats,showkeys,preprintnumbers,superscriptaddress,aps,prb]{revtex4-1}

\usepackage{graphicx}
\usepackage{float}
\usepackage{color}
\usepackage{bm}

\begin{document}

\preprint{Wu et al, Elastic properties of teragonal Heusler compounds.}

\title{A critical study of the elastic properties and stability of Heusler compounds: \newline
       Phase change and tetragonal $X_{2}YZ$ compounds}

\author{Shu-Chun Wu}
\affiliation{Max Planck Institute for Chemical Physics of Solids, D-01187 Dresden, Germany}

\author{S.~Shahab Naghavi}
\affiliation{Department of Materials Science and Engineering, Northwestern University, Evanston, Illinois 60208, USA} 

\author{Gerhard~H. Fecher}
\email{fecher@cpfs.mpg.de}
\affiliation{Max Planck Institute for Chemical Physics of Solids, D-01187 Dresden, Germany}

\author{Claudia Felser} 
\affiliation{Max Planck Institute for Chemical Physics of Solids, D-01187 Dresden, Germany}

\date{\today}

\begin{abstract}

In the present work, the elastic constants and derived properties of tetragonal
and cubic Heusler compounds were calculated using the high accuracy of the 
full-potential linearized augmented plane wave (FPLAPW). To find the criteria
required for an accurate calculation, the consequences of increasing the numbers
of $k$-points and plane waves on the convergence of the calculated elastic
constants were explored. Once accurate elastic constants were calculated,
elastic anisotropies, sound velocities, Debye temperatures, malleability, and
other measurable physical properties were determined for the studied systems.
The elastic properties suggested metallic bonding with intermediate
malleability, between brittle and ductile, for the studied Heusler compounds. To
address the effect of off-stoichiometry on the mechanical properties, the
virtual crystal approximation (VCA) was used to calculate the elastic constants.
The results indicated that an extreme correlation exists between the anisotropy
ratio and the stoichiometry of the Heusler compounds, especially in the case of
Ni$_{2}$MnGa.

Metastable, cubic Ni$_2$MnGa exibits a very high anisotropy ($\approx 28$) and
hypothetical, cubic Rh$_2$FeSn violates the Born-Huang stability criteria in the
$L2_1$ structure. The bulk moduli of the investigated tetragonal compounds do
not vary much ($\approx 130\ldots 190$~GPa). The averaged values of the other
elastic moduli are also rather similar, however, rather large differences are
found for the elastic anisotropies of the compounds. These are reflected in very
different spatial distributions of Young's moduli when comparing the different
compounds. The slowness surfaces of the compounds also differ considerably even
though the average sound velocities are in the same order of magnitude
($3.2\ldots 3.6$~km/s). The results demonstrate the importance of the elastic
properties not only for purely tetragonal Heusler compounds but also for phase
change materials that exhibit magnetic shape memory or magnetocaloric effects.

\end{abstract}

\keywords{Elastic constants, tetragonal Heusler compounds, cubic instability, 
          phase transition, VCA}
\maketitle

\section{Introduction}

Heusler-type intermetallic compounds $X_2YZ$ ($X,Y=$~transition metals, and
$Z=$~main group elements) have become of particular interest due to their
fascinating thermal, electrical, magnetic, and transport
properties~\cite{gfp11}. The Heusler compounds crystallize in a face-centered
cubic (fcc) lattice. They are distinguished in two groups: regular or inverse
Heusler compounds. Regular Heusler compounds belong to $Fm\overline{3}m$ (space
group no.~225) symmetry, and inverse Heusler compounds belong to
$F\overline{4}3m$ (space group no.~216). Both cubic phases may undergo a 
cubic--tetragonal phase transition, in which the regular Heusler compounds transform
from $Fm\overline{3}m$ to the tetragonal $I4/mmm$ (no.~139), and the inverse
Heusler compounds transform from $F\overline{4}3m$ to tetragonal
$I\overline{4}m2$ (no.~119). Thus, the parent cubic and obtained tetragonally
distorted phases obey a supergroup--subgroup relation.

Due to a simple feature of Heusler compounds, it is critically important to have
an instrument for phase prediction. For example, cubic ferromagnetic Heusler
compounds follow the Slater--Pauling rule for localized moment systems. Their
magnetic moment $m$ depends simply on the valence electron concentration $n_v$
with $m=n_v-24$. Further, prospective candidates for superconductivity include
certain Heusler compounds with 27 electrons that exhibit a saddle point at the
$L$ point close to $E_{\rm F}$ in the band structure according to the van Hove
scenario~\citep{WJG09,VH53}. On the other hand, a high density of states at the
Fermi level causes instability and a phase transition to lower symmetry forced
by a band Jahn--Teller distortion~\cite{BBC99,BOF11}. This
competition is one example that shows the importance of phase prediction in the
Heusler compounds. However, both tetragonal and cubic phases have their own
importance for industrial as well as fundamental research.

Tetragonally distorted Heusler compounds have attracted interest in the field of
spintronics, in particular, for spin-torque applications, owing to their
magnetic anisotropy in the perpendicular
axes~\cite{WBFFSBM08,KK70,WMWNOYT09,WMWSNO10}. Therefore, the theoretical
prediction of new materials with suitable design properties is active research
in this field~\cite{GDr10}.

In fact, processing and designing new materials requires knowledge of physical
properties, such as hardness, elastic constants, melting point, and ductility.
The calculation of elastic constants is an efficient and fast tool used for
elucidating physical properties as well as the mechanical stability and possible
phase transitions of crystalline systems. Applied strains, such as shear or
elongation, provide not only valuable information about the instability itself
but also the directional dependence of instabilities in crystals. The
directional dependence of instabilities becomes important when not only 
cubic--tetragonal but also cubic--hexagonal, tetragonal--hexagonal, and lower symmetry
phase transitions are relevant, such as those observed in
Mn$_3$Ga~\cite{WBFFSBM08,KK70}. Unlike mechanical instability, the determination
of elastic constants is essential for applications of magnetic shape memory
alloys. The elastic constants also provide valuable information on the
structural stability, anisotropic character, and chemical bonding of a
system~\cite{Gil09,Gil01,Pear97}. Moreover, other measurable properties can be
estimated using the elastic constants, such as the velocity of sound, Debye
temperature, melting point, and hardness. This information is an essential
requirement for both industrial applications and fundamental research. For
example, these properties are essential for studying superconductivity and heavy
fermion systems in which a drastic change of elastic constants and related
properties have been reported upon the phase transition~\cite{BWF94,LKG89}. The
elastic properties are so important that Gilman~\cite{Gil60} concluded: 
{\it ``the most important properties of a crystal are its elastic constants"}.

In the present study, some well-known tetragonal and cubic Heusler compounds are
examined and compared with the available experimental and theoretical
data~\cite{LZS13,LLH12,LHL12,LLH11,LLH11-doped,HLY09,MMP06,BRC03,JOW12}.
Starting from the cubic phase, cubic $L2_1$ Ni$_2$MnGa and Rh$_{2}$FeSn are
considered for detailed studies. For the tetragonally distorted systems,
Ni$_2$MnGa (in the non-modulated tetragonal ($c/a>1$) structure), Mn$_2$NiGa,
Fe$_2$MnGa, and Mn$_2$FeGa Heusler compounds are examined. The intermetallics
Mn$_{2}Y$Ga ($Y=$~Fe, Ni) and $X_{2}$MnGa ($X=$~Fe, Ni) undergo tetragonal
magneto--structural transitions that may result in half metallicity and magnetic
shape memory or magnetocaloric effects. In the case of Ni$_2$MnGa, the
composition dependence (chemical disorder effect) of the phase transition is
studied using the virtual crystal approximation (VCA). Calculating the
mechanical and elastic properties of off-stoichiometric compounds in the
tetragonal phases elucidates phase transformations. Elastic constants and
mechanical properties of some Rh-based Heusler compounds, reported by
Suits~\cite{Sui76}, are calculated. The dependence of the elastic constants and
the number of used $k$-points and plane waves (defined in full-potential
linearized augmented plane wave (FPLAPW) by $R_{\rm MT}k_{\rm max}$, where
$R_{\rm MT}$ is the muffin tin radius and $k_{\rm max}$ is the largest $k$
vector) are discussed in detail. The importance of using sufficiently large
numbers of $k$-points and plane waves for a reliable estimation of the elastic
properties is demonstrated.

The present work concentrates on the elastic properties of metastable cubic,
tetragonal, and phase change materials that exhibit magnetic shape memory or
magnetocaloric effects. The results for the famous series of half-metallic
Co$_2$-based Heusler compounds that has a high impact on magnetoelectronics will
be published elsewhere~\cite{WNF17}.

\section{Methodology}
\label{cal}

\subsection{Computational details}

In this section, the basic equations for calculating the elastic constants are
presented (for more details see Appendix~\ref{app:basic}). The most easily
determined quantity is the bulk modulus $B$, which provides the behavior of the
crystal volume or lattice parameters under hydrostatic pressure. There are
several ways to calculate the bulk modulus from the energy--volume 
$E_{\rm tot}(V)$ relation (see References~\onlinecite{ZSc03,Bri47,Mur44}). In the
present work, the bulk modulus $B$ is determined by fitting the total energy
calculations to the Birch--Murnaghan equation of state~\cite{Bri47,Mur44}.
According to this model, the dependence of the energy on the change in the
crystal volume $V$ under hydrostatic pressure $p$ is given by

\begin{equation}
   E = E_0 + \frac{9}{16} \frac{B V_0}{14703.6} \left[(6-4v^2)\eta^2 + B' \eta^3 \right],
\label{eq:eos}
\end{equation}

where $B'=dB/dp$ is the pressure derivative of the bulk modulus and $\eta=v^2-1$, 
with the ratio $v = (V_0/V)^{1/3}$ of the actual volume $V$ under pressure to 
the relaxed volume $V_0$ at the lowest total energy $E_0$ (see 
Reference~\onlinecite{Wall72}). The related pressure is given by

\begin{equation} 
   p = \frac{3}{2} B (v^7 - v^5) \left[1 + \frac{3}{4}(B'-4)\eta \right].
\end{equation}

For tetragonal crystals, the dependence of the bulk modulus $B$ on the elastic
stiffness is given by

\begin{equation} 
   B = \frac{1}{9}(2c_{11}+2c_{12}+4c_{13}+c_{33}). 
\end{equation}

In the case of cubic crystals, $c_{12}=c_{13}$, $c_{11}=c_{33}$, and
$c_{44}=c_{66}$ (see also Appendix~\ref{app:basic}). Therefore, the equations of
the elastic constants for cubic systems are easily obtained from the tetragonal
equations.

The remaining elastic properties are determined by applying different
types of strain ($e_i$) to the tetragonal lattice and by applying proper
relations between the total energy and the strain components. The energy
$E(e_i)$ of the strained lattice is calculated using Hooke's law (see
equation~(\ref{eq:hook})). According to Wallace~\cite{Wall72}, if the strains
$e_i$ are small, the change of the energy is given by 

\begin{equation}
  \frac{\Delta E}{V_0} = \sum \Delta_i e_i + \frac{1}{2} \sum \sum c_{i,j} e_i e_j.
\label{eq:strainenergy}
\end{equation}

Here, $\Delta E = E(e_i) - E_0$ with equilibrium energy $E_0$ at volume $V_0$
without strain. The linear terms vanish at equilibrium or if the strain causes
no change in the volume of the crystal. The elastic constants $c_{i,j}$ are
obtained from the second-order terms and are calculated from the second
derivatives of the energy with respect to the strains:

\begin{equation}
  c_{i,j} =  2 \left. \frac{\partial^2 E}{\partial e_i \partial e_j} \right|_{V=V_0}
\label{eq:dede}
\end{equation}

The second derivative relation of elastic constants ($c_{ij}$) with total energy
highlights the importance of an accurate calculation of the total energy.
Therefore, the choice of the density functional theory solver in the calculation
of elastic constants and related properties is significant.

For cubic crystals, three independent elastic constants need two
strains for calculations, while in tetragonal systems (space group Nos. 89--142),
six independent elastic constants need five different strains. In fact, there are
numerous ways to apply the six different strains and their combinations to the
crystal. The necessary side condition of equation~(\ref{eq:dede}) is that the
volume must be conserved when applying the strain. Therefore, the use of linear
strain components ($e_i=\delta$ in all possible cases and combinations in the
strain matrix of equation~\ref{eq:strain} of Appendix~\ref{app:basic}) would
lead to large uncertainties because they are not always volume conservative, or
they make the use of additional derivatives necessary (for example,
$\frac{\partial E}{\partial V}$, $\frac{\partial V}{\partial e_i}$, and higher
orders).

\begin{table*}[htb]
\centering
\caption{Strain table for calculation of the elastic constants in tetragonal systems. \newline
         Note that only types (1) to (5) are volume conservative. Only components with $e_i\neq0$
         are given. Type (0) corresponds to calculation of the bulk modulus. }
\begin{ruledtabular}
\begin{tabular}{llllll}
Type &  & Strain &  &  & $\Delta E / V_{0}$ \\
\hline
(0)  & isotropic    & $e_{1}=\delta$                    & $e_{2}=\delta$       & $e_{3}=\delta$                               & (see bulk modulus)  \\
(1)  & monoclinic   & $e_{1}=\delta^{2}/(1-\delta^{2})$ & $e_{4}=\delta$       &                                              & 2$c_{44}\delta^{2}+O(\delta^{4})$  \\
(2)  & triclinic    & $e_{3}=\delta^{2}/(1-\delta^{2})$ & $e_{6}=\delta$       &                                              & 2$c_{66}\delta^{2}+O(\delta^{4})$ \\
(3)  & orthorhombic & $e_{1}=\delta$                    & $e_{2}=-\delta$      & $e_{3}=\delta^{2}/(1-\delta^{2})$            & $(c_{11}-c_{12})\delta^{2}+O(\delta^{4})$ \\
(4)  & orthorhombic & $e_{1}=\delta$                    & $e_{2}=\delta^{2}/(1-\delta^{2})$ & $e_{3}=-\delta$                 & $(c_{11}-2c_{13}+c_{33})\delta^{2}/2+O(\delta^{4})$ \\
(5)  & tetragonal   & $e_{1}=\delta$                    & $e_{2}=\delta$       & $e_{3}=-\delta(2+\delta)/(1+\delta)^{2}$     & $(c_{11}+c_{12}+2c_{33}-4c_{13})\delta^{2}/2+O(\delta^{4})$ \\
(6)  & tetragonal   & $e_{3}=\delta$                    &                      &                                              & $c_{33}\delta^{2}/2$ \\
\end{tabular}
\end{ruledtabular}
\label{tab:strain}
\end{table*}

\begin{figure}[htb]
\centering
\includegraphics[width=8.5cm]{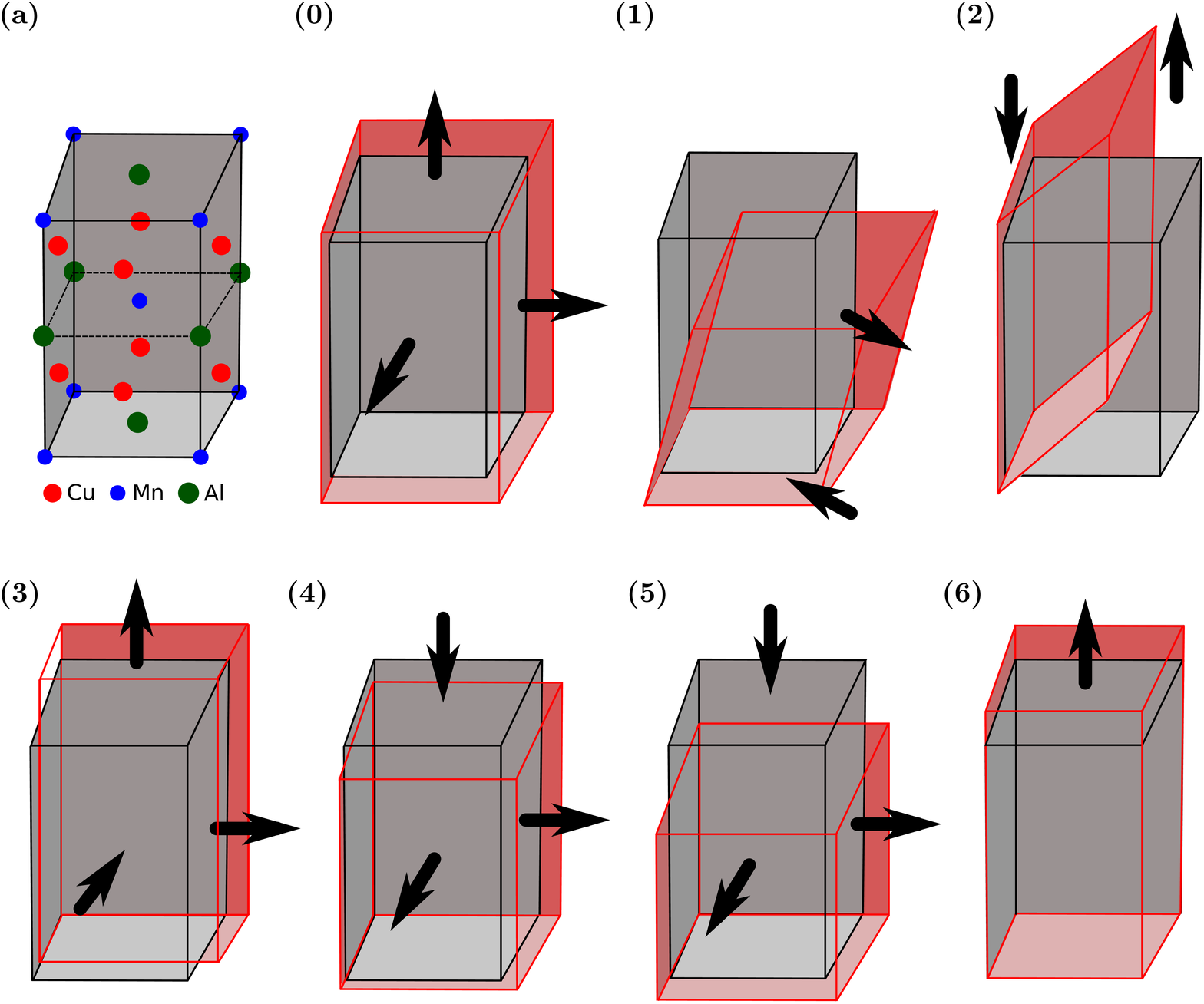}
 \caption{(Online color) Strain types for calculation of the elastic constants in tetragonal systems. \newline 
          (a) shows the tetragonal Heusler structure with $I\:4/mmmm$ symmetry.
          (0)--(6) show the strain types and resulting distortions according to Table~\ref{tab:strain}. }
\label{fig:straintypes}
\end{figure}

Table~\ref{tab:strain} and Figure~\ref{fig:straintypes} summarize the applied
strains that are used to determine the elastic constants in the present work.
The applied strains are the same as those reported by Kart {\it et al}~\cite{OUKC08}. 
The isotropic strain (0) is not used directly for the calculation of the elastic
constants, as it gives the same information as discussed for the bulk modulus
$B$ (see discussion above). The five strain types (Equations~(1)--(5) in
Table~\ref{tab:strain}) are chosen to be volume conservative. The last strain type does
not conserve the volume, but it keeps the same symmetry as the crystal and thus
can be calculated from the energy versus $c/a$ relation~\cite{OUKC08}, where
$c/a$ is the ratio of the two independent lattice parameters of the tetragonal
crystals.

In the present work, six distortions of each type in the range of $-3\% \leq
\delta \leq +3\%$ were applied to the relaxed structure with $V_0$ from the
structural optimization using the Birch--Murnaghan equation of state. For
tetragonal systems, the energy $E(\delta)$ versus applied strain curves were
fitted to a fourth-order polynomial 
$E(\delta)=E_0 + a_2 \delta^2 + a_3 \delta^3 + a_4 \delta^4$.

Here, an additional method of verifying the convergence as well as the accuracy
of the results is introduced. In principle, it is sufficient to use either
Equation~(4) or~(5) of Table~\ref{tab:strain} to calculate all six elastic
constants. However, the elastic constants and all related properties are
calculated with both equations to ensure that they provide the same results.
This happens, indeed, only if the results are well converged (see also
Section~\ref{Cubic}). The system is overdetermined by using both types of
strains; however, in this way, the accuracy of the calculated quantities can be
estimated. In fact, the values reported here have an error below 0.5\%. The
combination of different strains as well as different types of equations of
state allow determination of the uncertainty of the calculated results, which is
expected for a reliable computer experiment.
 
\subsection{Electronic structure calculations}
\label{opt}

The {\it ab-initio} electronic structure calculations were performed using the
\textsc{Wien2k} code~\cite{Wien2k}. The all-electron full-potential
method, FLAPW, with an unbiased basis covers all elements of the periodic table
with any spin configuration. This feature is essential for Heusler compounds
because they may contain diverse types of atoms, including lanthanide and
actinide atoms, together with exotic magnetic ordering. The accuracy of this method
makes it suitable for the studied systems. For example, Co$_2$TiAl fails
with spherical potentials or full symmetry potentials together with bare
exchange--correlation functionals neglecting gradient
corrections~\cite{GFBWF09,IAKI82,MBS95}. Since elastic constants are calculated
from the second derivatives of the total energy, an accurate calculation of total
energy is extremely important.

The exchange--correlation functional was taken in the generalized gradient
approximation of Perdev, Burke, and Enzerhoff (GGA-PBE)~\cite{PBE2,PBE0-2}. The
number of plane waves was restricted by $R_{\rm MT}k_{\rm max}=9$, and the number
of $k$-points was set to 8000 $k$-points in the full Brillouin zone. As
discussed in Section~\ref{Cubic}, these criteria ensure the convergence of the
calculated elastic properties for the investigated systems.

The lattice parameters were optimized before calculating the elastic constants.
The results of the structural optimizations are summarized in
Table~\ref{tab:latt1} along with some previously reported experimental and
theoretical values. Here, the $c/a$ ratio was obtained by a full optimization of
the Heusler compounds in tetragonal space groups 119 or 139. In other words, to
find the energy minimum, not only the $c/a$ ratio changed (the elongation of
$c$) ~\cite{OCc10,OUKC08} (see also Figure~\ref{fig:Ni2MnGa}) but also the
volume of the structures was relaxed. The assignment of lattice parameters
should be performed carefully since the optimization was performed in the
tetragonal symmetry. When reducing the cubic $fcc$ cell to a tetragonal $fct$
cell, the cubic lattice parameter $a=a_c$ becomes $c$, and the tetragonal
parameter $a=a_t$ becomes $a_c/sqrt(2)$. To better understand the distortion of
the cubic Heusler structure to the tetragonal Heusler structure, the distortion
parameter $\varepsilon$ is defined by

\begin{equation}
  \varepsilon =  \frac{c}{a_t} \frac{1}{\sqrt{2}}-1,
\label{eq:distortion}
\end{equation}

where $a_t$ is the tetragonal $a$ parameter. At $\varepsilon=0$, the structure is
cubic; at $\varepsilon<0$, the cubic cell is compressed; and at $\varepsilon>0$,
the cubic cell is elongated along one of the principal axes.

\begin{table}[htb]
\centering
\caption{Results of the structural optimization. Lattice parameters $a$ and $c$ are given in 
        {\AA}, $\varepsilon$ is dimensionless, and total magnetic moments $m$ are in $\mu_{\rm B}$.
         All structures are fully optimized ($V$, $c/a$, magnetic state) for the given symmetries. 
         Column {\bf sym} gives the number of the corresponding space group, and cubic and tetragonal 
         structures are assigned by the Pearson symbols $cF16$ and $tI4$, respectively.
         Experimental values are given for comparison. Note that the cubic variant of Rh$_2$FeSn
         is only hypothetical. }
  
  \begin{ruledtabular}
  \begin{tabular}{llcccc}
   Compound  & {\bf sym} & $a_c$, $c$ & $a_t$ & $\varepsilon$ & $m_{\rm tot}$ \\
  \hline   
   Mn$_2$NiGa                 & $tI4$ 119  & 6.91 & 3.78 & 0.293 & 1.0  \\
  \hline                                                   
   Ni$_2$MnGa                 & $tI4$ 139  & 6.80 & 3.78 & 0.272 & 4.0  \\
   {\it Exp.}\footnotemark[1] &            & 6.44 & 3.90 & 0.168 & 4.09 \\
  \hline                                                   
   Ni$_2$MnGa                 & $cF16$ 225 & 5.81 &      & 0     & 4.1  \\
   {\it Exp.}\footnotemark[3] &            & 5.82 &      & 0     & 4.17 \\
  \hline                                                   
   Mn$_2$FeGa                 & $tI4$ 119  & 7.30 & 3.68 & 0.403 & 0.77 \\
  \hline                                                   
   Fe$_2$MnGa                 & $tI4$ 139  & 7.31 & 3.62 & 0.428 & 0.13 \\
  \hline                                                   
   Rh$_2$CrSn                 & $tI4$ 139  & 7.33 & 4.08 & 0.270 & 2.40 \\
   {\it Exp.}\footnotemark[2] &            & 7.16 & 4.09 & 0.238 &      \\
  \hline                                                   
   Rh$_2$FeSn                 & $tI4$ 139  & 7.12 & 4.13 & 0.219 & 3.92 \\
   {\it Exp.}\footnotemark[2] &            & 6.91 & 4.15 & 0.177 & 3.70 \\
  \hline                                                   
   Rh$_2$FeSn                 & $cF16$ 225 & 6.25 &      & 0     & 3.54 \\    
  \hline                                                   
   Rh$_2$CoSn                 & $tI4$ 139  & 7.22 & 4.05 & 0.261 & 2.25 \\
   {\it Exp.}\footnotemark[2] &            & 6.90 & 4.14 & 0.179 & 2.44 \\
  \end{tabular}
  \end{ruledtabular}
\footnotetext[1]{Reference~\onlinecite{MKo92}}
\footnotetext[2]{Reference~\onlinecite{Sui76}}
\footnotetext[3]{Reference~\onlinecite{WZT84}}
  \label{tab:latt1}
  \end{table}

The magnetic state was verified using different settings of the initial
magnetization: ferromagnetic (all initial spins parallel) or ferrimagnetic
(initial spins partially antiparallel). From the studied systems, all Mn$_2$YZ
compounds exhibit ferrimagnetic order in which one Mn has majority and the other
Mn has minority orientation. For all other compounds, the ferromagnetic ground state
has the lowest total energy.

\section{Results}

Prior to discussing the tetragonal Heusler compounds, the elastic constants of
the unstable and metastable cubic systems are discussed. Here, Ni$_2$MnGa with
$L2_{1}$ structure is selected as a metastable system. This compound is one of
the most investigated materials owing to its shape memory behavior and its
potential applications in actuator devices. In fact, in Section~\ref{Cubic},
only the cubic phase is addressed, and the tetragonal phase of Ni$_2$MnGa is
discussed in Section~\ref{Tetra}. Moreover, Ni$_2$MnGa was used as an example
case to study the significance of increasing the number of $k$-points and plane
waves and their relations to the convergence of the elastic constants. In the
second part of this section, the tetragonal phase of Heusler compounds are
discussed. The elastic constants together with the corresponding measurable
properties for selected tetragonal Heusler compounds are investigated. The role
of the stoichiometry on the phase transition of Ni$_2$MnGa is also explored.

\subsection{Elastic constants and metastability in cubic and tetragonal compounds.}
\label{Cubic}

\begin{figure}[htb]
\centering
\includegraphics[width=8.5cm]{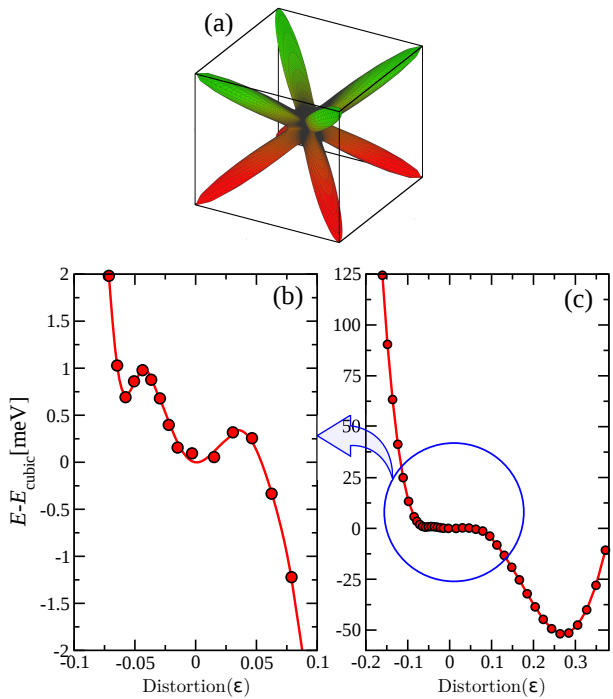}
 \caption{(Online color) (a) The calculated spatial distributions of the rigidity modulus $G(\hat{r})$ 
           of Ni$_2$MnGa. (b, c) The total energy as a function of tetragonal strain 
           along the $c$-axis at constant volume. Panel (b) focuses on the 5M modulated phase with $c/a\approx0.9$}
\label{fig:Ni2MnGa}
\end{figure}

Stoichiometric Ni$_2$MnGa undergoes a structural phase transition from the
austenite into the martensite phase~\cite{TSOTA08}. Depending mostly on the
composition, the martensite structure is characterized by the tetragonal $5M$
modulated structure with $c/a\approx0.94$, the orthorhombic $7M$ structure with
$c/a\approx0.9$, and the non-modulated tetragonal structure with
$c/a\approx1.2$~[\onlinecite{OCc10}]. Figure~\ref{fig:Ni2MnGa} shows the
appearance of different stable and metastable phases with varying $c/a$
elongation. To focus on the cubic phase, Figure~\ref{fig:Ni2MnGa}(b) shows only
the small range of strains with $c/a<1$ that cover the cubic
phase. The deepest energy minimum is located at a strain of about
$\varepsilon=0.27$, corresponding to $c/a \approx1.26$ (non-modulated phase). A shallow
minimum (see Figure~\ref{fig:Ni2MnGa}(b)) appears at a strain of about -0.05
($c/a \approx0.94$, 5M phase). The elastic constants of the structure with
$c/a>1$ will be discussed in Section~\ref{Tetra}.

\begin{figure}[htb]
\centering
\includegraphics[width=8.5cm]{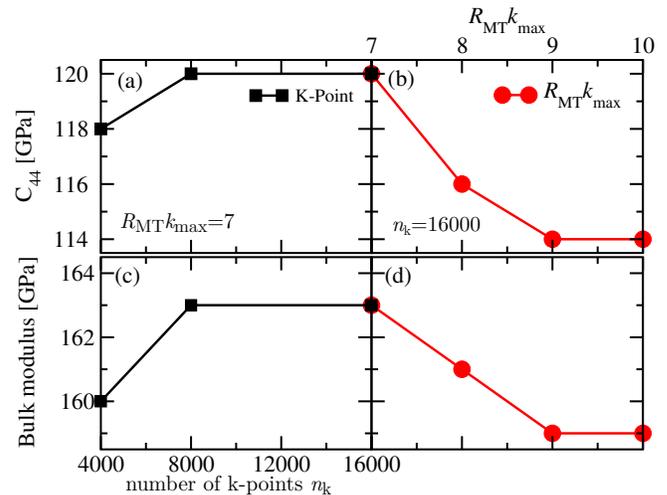}
   \caption{(Online color) Convergence of elastic properties. \newline
            The shear modulus $c_{44}$ and bulk modulus ($B$) as a function 
            of $R_{\rm MT}k_{\rm max}$ and the number of $k$-points. The right side panels 
            (a,c) show the increment of $B$ and $c_{44}$ as a function of the $k$-points
            with $R_{\rm MT}k_{\rm max}$=7. Using the converged 
            $k$-points (16000 or 8000), in the left side panels (b,d), these values 
            decrease with increasing $R_{\rm MT}k_{\rm max}$. Therefore,
            the minimum number of $k$-points and value of $R_{\rm MT}k_{\rm max}$
            are 8000 and 9, respectively.}
\label{fig:criteria}
\end{figure}

As shown in Figure~\ref{fig:Ni2MnGa}(c), the metastable cubic phase exists only
under an infinitesimal strain and exhibits a very low energy modulation. The
optimized lattice constant of cubic Ni$_2$MnGa with the $L2_1$ structure is
5.81~{\AA}, in excellent agreement with the experimental value of
5.82~{\AA}~[\onlinecite{WZT84}]. This phase is only stable within
$\pm1$~meV energy changes, which confines the lattice distortion to $<\pm1$\%.
This distortion relates to the tetragonal distortion, providing the 
$c_{11}-c_{12}$ combination of elastic constants. To observe such a non-trivial change
in energy and to have a smooth dependence on strain, the results need to be
precisely converged with very high precision. This does not imply, however, that
the results do not need to be converged for a wide energy window with a deep
minimum. As an example, the importance of converged results is demonstrated in
Figure~\ref{fig:criteria}. The $c_{44}$ shear modulus is stable for a lattice
distortion of about $\pm3$\% and an energy change of more than $\pm30$~meV. In
this case, the rough calculation provides a smooth curve, but the calculated
elastic constants significantly deviate from the converged results. The
convergence of the results have the same importance for the calculation of the
bulk modulus ($B$) (see Figure~\ref{fig:criteria}(c,d)).

Figure~\ref{fig:criteria} shows the convergence of $c_{44}$ and $B$ with respect
to $R_{\rm MT}k_{\rm max}$ and number of $k$-points. As shown in (a,c) for
constant $R_{\rm MT}k_{\rm max}=7$, increasing the number of $k$-points
increases the $c_{44}$ and $B$ values. These results converge at 8000 
$k$-points. In contrast, increasing $R_{\rm MT}k_{\rm max}$ decreases $c_{44}$ and
$B$. Note that a similar result could be obtained at a more relaxed criterion,
for example at 2000 $k$-points and $R_{\rm MT}k_{\rm max}=7$, due to error
cancellations. Therefore, $R_{\rm MT}k_{\rm max}=9$  and 8000 $k$-points are the
minimum criteria to converge the results for the systems studied in this work.

\begin{table}[htb]
\centering
\caption{Elastic properties of metastable Ni$_2$MnGa and hypothetical Rh$_2$FeSn
         with $L2_1$ structure. \newline
         The calculated elastic parameters $c_{ij}$, $c'$, and $B$ are given in GPa, and
         the cubic elastic anisotropy $A_{e}= \frac{2 c_{44}}{c_{11} - c_{12}}$ is dimensionless
         and corresponds to the tetragonal $A_{001}$.}

  \begin{ruledtabular}
  \begin{tabular}{lcccccc}
    Compounds & $c_{11}$ & $c_{12}$ & $c_{44}$ & $c'$ & $B$ & $A_{e}$ \\
  \hline                                                                
   Ni$_2$MnGa                        & 164 & 156 & 115  & 4.1 & 159 & 28.02   \\
   {\it Exp.}\footnotemark[1]        & 152 & 143 & 103  & 4.5 & 146 &         \\
   {\it Other calc.}\footnotemark[2] & 163 & 152 & 107  & 5.5 & 156 &         \\
   Rh$_2$FeSn                        & 124 & 206 & 84.3 & -82 & 179 & $<0$   \\
  \end{tabular}  
  \end{ruledtabular}
  
  \footnotetext[1]{Reference~\onlinecite{WPT96}}
  \footnotetext[2]{Reference~\onlinecite{OCc10}}
  \label{tab:elast1}
\end{table}

The calculated elastic constants of Ni$_2$MnGa with $L2_1$ structure are given
in Table~\ref{tab:elast1}. The calculated results show a reasonable agreement
with the experimental results and coincide with previously reported theoretical
results~\cite{OUKC08,OCc10}. Note, however, that good agreement with the
experiment results does not guarantee the accuracy of the calculations. First,
the experiments were performed at 300~K for Ni$_2$MnGa in the $L2_1$ phase, and
off-stoichiometry has a significant effect on the measured elastic
constants~\cite{QCRSBDBL09}. Moreover, the employed experimental method may
result in different measured elastic constants~\cite{WPT96}. As an example, the
$c_{44}$ value deviates by about 60~GPa based on the experimental method. In
general, the measured elastic constants are inversely related to
temperature~\cite{Led06}. Hence, a higher value should be
expected for the calculations. Fortunately, $c'$, which is a difference between
two constants ($c_{11}$ and $c_{12}$), is argued to be less dependent on
temperature~\cite{Led06}. In fact, the calculated $c'=4.5$~GPa
exhibits a better agreement with the experiment (4.1~GPa) than previously
reported values~\cite{OUKC08,OCc10}.

In addition, $c_{12}$ and $c_{11}$ do not have any physical basis; in other
words, no phonon mode directly corresponds to these constants. Mixing with other
stiffness ($c_{ij}$), however, results in a meaningful combination. For example,
the tetragonal shear modulus $c'=(c_{11}-c_{12})/2$ corresponds to cubic--tetragonal
distortion. Moreover, it is well established that $c'$ --~associated with slow
transverse acoustic waves~\cite{WPT96}~-- plays an important role in the
occurrence of structural transformations. Another important quantity is the
Cauchy pressure $c_{p}=c_{12}-c_{44}$. A negative value of $c_{p}$
($c_{12}<c_{14}$) may indicate covalent bonds, where the angular dependence of
the inter-atomic forces becomes important.

Furthermore, detailed analysis of elastic constants sheds light on the stability
and phase transition in Heusler compounds. Cubic Ni$_2$MnGa with soft $c'$ is on
the border of the phase transition. With the similar interpenetration 
(see Appendix~\ref{app:stability}), the large elastic
anisotropy $A_{e}$ of Ni$_2$MnGa ($A_{e}$=28) hints on its tendency to deviate
from the cubic structure. Anisotropy is another indicator for the instability of cubic
structures. The elastic anisotropy of crystals is also an important parameter
for engineering since it correlates to the possibility of micro-cracks in
materials. Unlike the mechanical properties, anisotropy shows the
tendency of a system toward phase transitions as it inversely relates to the $c'$
parameter. In fact, an illustrative way to show the anisotropy is to visualize
the rigidity modulus $G(\hat{r})$ or Young's modulus $E(\hat{r})$.
Figure~\ref{fig:Ni2MnGa}(a) shows that the rigidity modulus is largest in the
{$\left<111\right>$}-type direction that is along the tetragonal axes. Such a
significant deviation from spherical shape indicates that the moduli of
Ni$_2$MnGa exhibit a large degree of anisotropy. In principle, when 
$c_{11}-c_{12}\rightarrow~0$, the rigidity distribution exhibits a stronger 
directional dependency, as shown for Ni$_2$MnGa in Figure~\ref{fig:Ni2MnGa}(a).

\begin{figure}[htb]
\centering
\includegraphics[width=8.5cm]{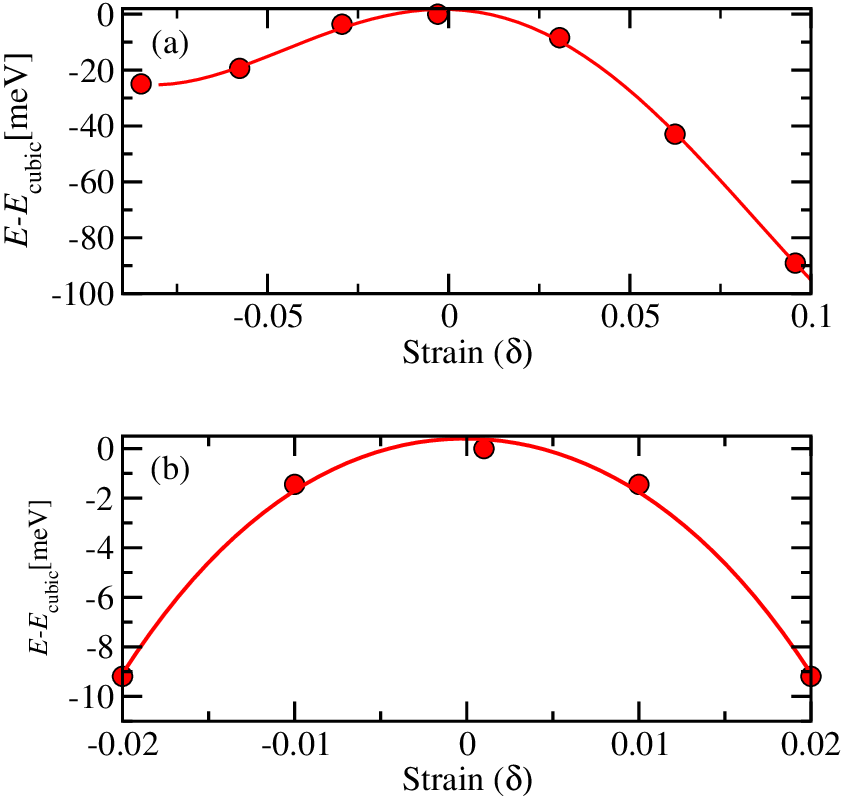}
 \caption{(Online color) Energy--strain relation for {\it hypothetical}
           Rh$_2$FeSn with $L2_1$ structure. \\
           Displayed are the calculated total energies as function of (a) tetragonal 
          (symmetry No.~139) and (b) orthorhombic (symmetry No.~69) strains. }
\label{fig:Rh2FeSn}
\end{figure}

In the next step, a compound that is not stable in the cubic structure was
examined for comparison. In the case of Rh$_2$FeSn shown in
Figure~\ref{fig:Rh2FeSn}, the cubic structure exhibits a maximum of the total
energy, and any tetragonal distortion will lead to a different stable structure.
Based on Figure~\ref{fig:Rh2FeSn}, the appearance of energy minima are expected
for two different tetragonally distorted systems with $c/a>1$ and $c/a<1$.
Larger distortions show that $c/a>1$ is the stable phase, while $c/a<1$ is a
metastable phase. Previous works only reported the structure with
$\varepsilon>0$~\cite{Sui76,Rh2FeSn-alt_HU}. However, the calculated elastic
constants also supported the instability of the cubic phase from negative values
of tetragonal shear modulus $c'$ and anisotropy $A_{\rm e}$. The metastable
structure with $\varepsilon<0$ appears when expanding the in-plane lattice
parameter $a$ while keeping $c$ at the cubic lattice parameter. Such a situation
may be artificially initialized by epitaxial or pseudomorphic thin film growth
on a substrate with an appropriate lattice parameter. Similar metastable
situations may exist in many other tetragonal Heusler compounds and will open
the field of lattice parameter engineering to enlarge the number of properties
on demand.

\subsection{Tetragonal Heusler compounds}
\label{Tetra}

The tetragonal Heusler compounds studied in this work along with their elastic
constants are summarized in Tables~\ref{tab:mech} and~\ref{tab:vel}. The Heusler
intermetallics Mn$_{2}Y$Ga ($Y=$Fe, Ni) and $X_{2}$MnGa ($X=$Fe, Ni) undergo
tetragonal magneto-structural transitions that result in half-metallicity,
magnetic shape memory, or magneto-electric effects. In this section, the 
off-stoichiometric compositions are briefly discussed, and then, the elastic
constants and related properties of the Heusler compounds are analyzed.
Calculating the elastic properties of the tetragonal phases illuminates the
structural transformations, chemical bonding, and mechanical stability of these
intermetallic compounds for applications. Likewise, the elastic properties of
Rh-based Heusler compounds synthesized by Suits~\cite{Sui76} are calculated.
Although Ni$_2$MnGa has been widely studied experimentally at different phases,
there is no experimental measurement of the elastic modulus of the non-modulated
tetragonal phase, and only theoretical works on this phase have already been
reported~\cite{OCc10,OUKC08}.

\begin{table*}[htb]
\centering
\caption{Elastic properties of selected tetragonal Heusler compounds.
         The different moduli ($B, G, E$) and elastic constants $c_{ij}$ are given in
         GPa, the corresponding elastic compliances $s_{ij}$ are in (TPa)$^{-1}$.
         The anisotropies ($A_{\rm B}$ , $A_{100}$, $A_{001}$),
         Phug's ratio $k$, and Poisson's ratio $\nu$ are dimensionless. 
         The compressibility $\kappa$ is in GPa$^{-1}$. For Ni$_2$MnGa,
         results taken from Reference~\onlinecite{OCc10} are given in brackets (). }

  \begin{ruledtabular}
  \begin{tabular}{ll|lllll|lllll}
               & Mn$_2$NiGa &                  && Ni$_2$MnGa  &&               & Mn$_2$FeGa & Fe$_2$MnGa & Rh$_2$CrSn & Rh$_2$FeSn & Rh$_2$CoSn \\
               &            & normal   & $^{25.1}$Mn & $^{30.9}$Ga & $^{28.1}$Ni & $^{27.9}$Ni &&&&&\\
  \hline
  $B_{\rm V}$  & 134.9  & 159.2 (158)  &  161.8  & 158.9   & 158.0   & 161.3   & 141.2   & 154.7  & 176.6  & 178.8  & 187.8 \\
  $B_{\rm R}$  & 126.5  & 159.2 (157.8)&  161.7  & 131.6   & 157.5   & 155.1   & 134.6   & 154.3  & 161.6  & 176.5  & 185.8 \\
  $B$          & 130.7  & 159.2 (157.9)&  161.8  & 145.2   & 157.8   & 158.2   & 137.9   & 154.5  & 169.1  & 177.7  & 186.8 \\
  $G_{\rm V}$  & 109.3  & 95.9  (73.8) &   95.9  &  96.0   &  93.7   &  95.1   & 106.6   & 116.2  & 108.3  & 86.2   & 101.5 \\
  $G_{\rm R}$  & 80.5   & 66.3  (53.8) &   66.6  &  57.2   &  62.5   &  50.3   & 37.9    & 56.1   & 85.2   & 74.6   & 76.5  \\
  $G$          & 94.8   & 81.1  (63.6) &   81.2  &  76.6   &  78.1   &  72.7   & 72.3    & 86.2   & 96.7   & 80.4   & 89.0  \\
  $E$          & 229.2  & 208   (208)  &  208.8  & 195.4   & 201.1   & 189.1   & 184.6   & 218.0  & 243.7  & 209.6  & 230.4 \\
  $c'$         & 67     & 60           &   62    &  60     & 58      &  67     & 12      & 21     & 73     & 70     & 67    \\
  \hline                                                                                          
  $c_{11}$     & 194    & 227 (252)    &  235    & 196     & 229     & 222     & 181     & 195    & 234    & 252    & 257    \\
  $c_{12}$     & 60     & 108 (74)     &  112    &  77     & 113     &  88     & 157     & 154    & 88     & 113    & 123    \\
  $c_{13}$     & 118    & 140 (144)    &  142    & 157     & 139     & 156     & 110     & 122    & 160    & 153    & 168    \\
  $c_{33}$     & 232    & 199 (194)    &  196    & 255     & 184     & 208     & 151     & 205    & 300    & 265    & 258    \\
  $c_{44}$     & 163    & 148 (100)    &  145    & 151     & 143     & 150     & 167     & 177    & 154    & 124    & 154    \\
  $c_{66}$     & 110    & 95  (55)     &  100    &  93     &  99     &  91     & 155     & 162    & 112    & 66     & 94     \\
                                                            
 \hline                                                                                           
  $s_{11}$     & 7.45   &  7.80  (7.26)&   7.61  & 10.43   &  8.05   & 10.21   & 23.26   & 14.76  & 6.72   & 6.25   &  6.82  \\
  $s_{12}$     & 0.02   & -0.58  (1.65)&  -0.57  &  2.05   & -0.53   &  2.76   & -17.60  & -9.83  & -0.14  & -0.94  & -0.66  \\
  $s_{13}$     & -3.80  & -5.09  (-6.61&  -5.09  & -7.69   & -5.66   & -9.69   & -4.13   & -2.95  & -3.51  & -3.06  & -3.99  \\
  $s_{33}$     & 8.17   &  12.21 (15.0)&  12.47  & 13.40   & 13.97   & 19.29   & 12.66   & 8.41   & 7.06   & 7.30   &  9.05  \\
  $s_{44}$     & 6.11   &  6.80  (10.0)&   6.89  &  6.63   &  7.01   &  6.67   & 6.00    & 5.66   & 6.47   & 8.06   &  6.47  \\
  $s_{66}$     & 9.01   &  10.51 (18.2)&  10.04  & 10.74   & 10.10   & 10.98   & 6.46    & 6.16   & 8.89   & 15.05  &  10.64 \\
  \hline                                                                                          
  $A_{\rm B}$  & 0.15   & 0.93         & 1.17    & -0.41   & 1.42    & -0.03   & 2.89    & 1.27   & 0.015  & 0.52   & 0.49   \\
  $A_{100}$    & 3.44   & 4.04         & 3.94    & 4.39    & 4.20    & 5.04    & 5.97    & 4.57   & 2.88   & 2.35   & 3.44   \\
  $A_{001}$    & 1.64   & 1.59         & 1.63    & 1.56    & 1.70    & 1.36    & 12.64   & 7.98   & 1.54   & 0.95   & 1.41   \\
  \hline                                                                                          
  $k$          & 1.38   & 1.96 (2.48)  & 1.99    & 1.90    & 2.02    & 2.18    & 1.91    & 1.79   & 1.75   & 2.20   & 2.09   \\
  $\nu$        & 0.21   & 0.282 (0.322)& 0.28    & 0.28    & 0.29    & 0.30    & 0.28    & 0.26   & 0.26   & 0.30   & 0.29   \\
  $\kappa$     & 0.0079 & 0.0063       & 0.0062  & 0.0076  & 0.0063  & 0.0064  & 0.0079  & 0.0065 & 0.0062 & 0.0056 & 0.0054 \\
  \end{tabular}
  \end{ruledtabular}
  \label{tab:mech}
  \end{table*}

As shown in Table~\ref{tab:mech}, in the case of Ni$_2$MnGa, the results show a
qualitative agreement with the previous theoretical report (values in brackets).
However, a quantitative comparison of the results reveals some significant
deviations. These differences can be traced back to the calculation method and
the method of performing structural optimization. To address this problem, the
cubic phase is briefly considered. In the cubic phase, the present as well as
other calculations are performed for the same lattice parameters using different
calculation schemes. Here, the results of FPLAPW calculations are slightly
larger than those of projected augmented wave (PAW) calculations, and the
deviations range from 1\% for $c_{11}$ up to 7\% for $c_{44}$. These small
differences are expected because of the selected methods and convergence
criteria. In contrast, in the case of the tetragonal system (see
Table~\ref{tab:mech}), the deviations between calculations range up to 30\%,
such as the case of $c_{44}$ and $c_{66}$. Indeed, the different calculation
methods should not lead to such a large discrepancy (if all factors are set
carefully), and these observed differences mainly arise from the underlying
structural optimization. Here, all initial tetragonal structures are fully
optimized at their relevant symmetries ($I4/mmm$ or $I\overline{4}m2$), and
thus, they differ from the simply elongated cubic structures (see also
Section~\ref{opt} for more details about the calculations).

\begin{figure}[htb]
\centering
   \includegraphics[width=8.5cm]{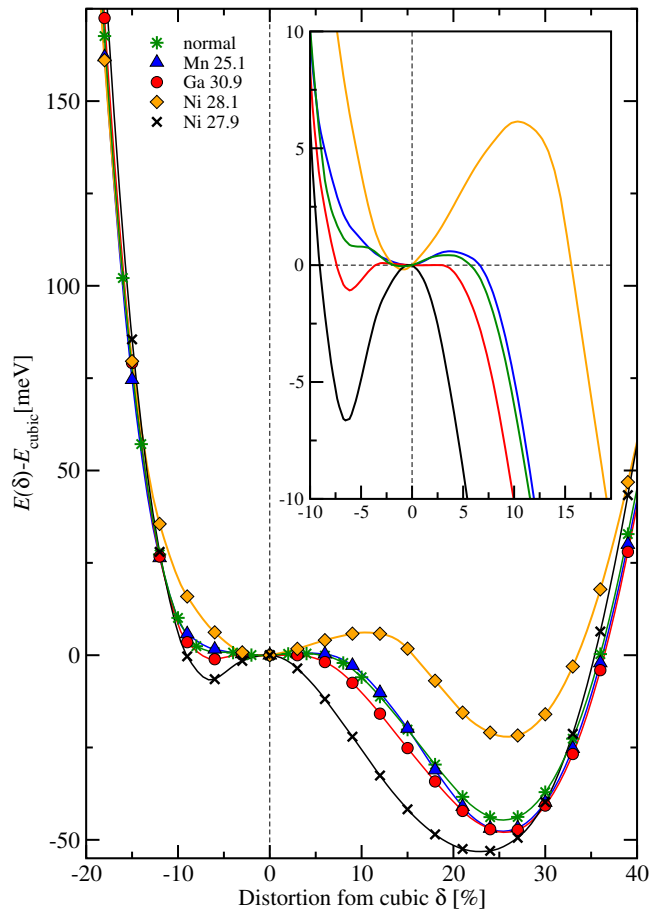}
   \caption{(Online color) The energy change along $c/a$ is plotted for the non-stoichiometry composition
   of Ni$_2$MnGa. The energy landscape (here, along the c-axis) is extremely sensitive to the stoichiometric composition, in 
   particular, the Ni component.
   }
\label{fig:vca}
\end{figure}

The elastic constants of all studied tetragonal Heusler compounds follow the
inequality $B>c_{44}>G>c'>0$, so that the tetragonal shear modulus $c'$ is the
main constraint on the stability and properties. Pugh's and Poisson's ratio (see
Appendix~\ref{app:basic}) supply valuable information about the malleability and
the type of bonding in crystals. Small values of Pugh's ratio indicate low
malleability of crystals~\cite{Pug54}, meaning they are brittle. Pugh's ratio
indicates the type of bonding, namely covalent or metallic bonding, because
changes in the angle of a covalent bond require more energy than stretching--
stressing the bond, meaning $G$ becomes larger compared to $B$; this leads to
smaller values of $k$. On the other hand, Poisson's ratio for covalent bonding
is about $\nu=0.1$ and increases for ionic crystals up to $\nu=0.25$. For
instance, Pugh's and Poisson's ratios of strong covalent compounds such as
diamond are assumed to be $k=0.83$ and $\nu=0.069$, respectively, while in the
strongly ionic KCl, Pugh's and Poisson's ratios are $k=1.19$ and $\nu=0.27$,
respectively~\cite{Diamond_HU,NaCl-KCl_HU}. Despite having different types of
bonding, both KCl and diamond are brittle. On the other hand, strongly malleable
gold exhibits ratios of $k=6.14$ and $\nu=0.42$~\cite{NCL12,Pug54}, and gold is
the most ductile metallic element. According to Christensen~\cite{Chr13}, the
critical value for the ductile--brittle transition appears at a Pugh's ratio of

\begin{equation}
	k_{B/D} = \frac{2}{3}\frac{1}{(1-\sqrt{1/2})} \approx 2.3,
\end{equation}

which corresponds to a critical Poisson's ratio of 
$\nu_{B/D}=(3\sqrt{2}-1)^{-1}\approx0.31$.
This criterion will be analyzed in more detail in a forthcoming work on cubic
Heusler compounds~\cite{WNF17}. Based on the calculated elastic
constants, the studied tetragonal compounds exhibit Pugh's ratios between 1.38
and 2.2 and Poisson's ratios between 0.21 and 0.3, indicating that they have an
intermediate behavior between ductile- and brittle- type behavior. The values of
$k$ and $\nu$ indicate the covalent or metallic character of these systems. More
details about the bonding type may be found from a Bader analysis of the charge
densities~\cite{Bad90,OBP09,OJL14}.

\begin{figure}[htb]
\centering
\includegraphics[width=8.5cm]{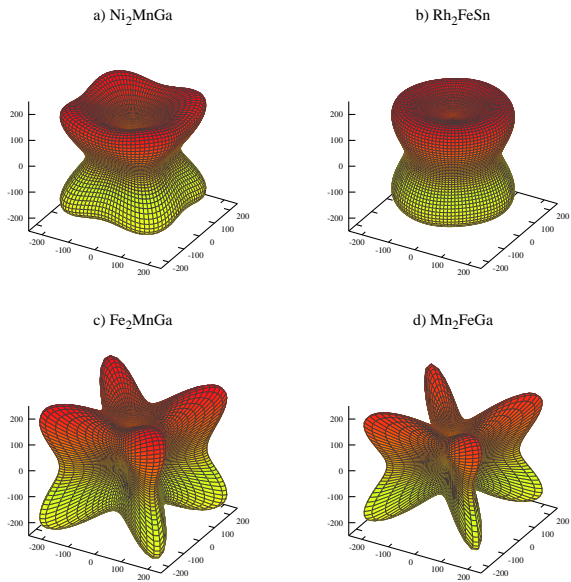}
   \caption{(Online color) Calculated spatial distribution of Young's moduli $E(\hat{r})$
            of some Mn- and Rh-based Heusler compounds. The change from the
            round shape of Rh$_2$FeSn to the sharply elongated shape in Mn$_2$FeGa indicates 
            an increase of the elastic anisotropy (see also Table~\ref{tab:mech}).}
\label{fig:bmym}
\end{figure}

The shear anisotropy factor $A_{\rm e}$ provides a measure of the degree of
anisotropy of the bonds between atoms in different planes. Tetragonal systems
are described by two different shear anisotropic factors $A_{001}$  and
$A_{100}$ (or equivalent $A_{010}$). Young's moduli of some of the studied
Heusler compounds are shown in Figure~\ref{fig:bmym}. The shear anisotropy for
\{100\} planes is considerably higher compared to \{001\} planes for Mn$_2$NiGa
and Rh$_2$FeSn. This behavior is similar for Mn$_2$FeGa and other Rh$_2$-based
compounds, as shown in Table~\ref{tab:mech}. In contrast, in the case of
Fe$_2$MnGa and Mn$_2$FeGa,  the anisotropy for \{001\} is higher than that for
\{100\}. Moreover, Rh$_2$FeSn has an interesting distribution of Young's
modulus: it is isotropic in the square x-y planes of the tetragonal structure,
arising from the value of $A_{001}=0.95$, which is close to unity.

\subsection{Virtual Crystal Approximation}

One interesting feature of Heusler compounds, including Ni$_2$MnGa, is their
sensitivity to stoichiometric compositions. An infinitesimal deviation may lead
to a phase transition or to changes in the electronic structure properties.
Here, VCA was applied to explore off-stoichiometric systems. This approximation
is valid for small changes of components with nearly the same radius, which
holds for the considered systems. Moreover, VCA is only valid for neighboring
elements and for small differences in the number of valence electrons 
($\Delta e^-<0.1$).

In VCA, the atom $^ZA$ with charge $Z$ is replaced by atom $^{Z'}A$ with the
virtual charge $Z'=Z\pm\epsilon$ to reflect that the average charge deviates
from the original value at a certain position. In Ni$_2$MnGa, if the site where
$^{28}$Ni resides is partially occupied by $^{25}$Mn, the charge at that site
will be lower; accordingly, the charge will be higher when $^{30}$Ga occupies
the same site. In parallel to the change in the nuclear charge, the number of
electrons in the primitive cell changes to remain neutral. Thus, for
$^{28\pm\epsilon}$Ni$_2$$^{25}$Mn$^{31}$Ga, the number of valence electrons will
be $n_v=30\pm2\epsilon$ in the primitive cell, whereas the number of core plus
semi-core electrons stays fixed at 72. The following cases are used in the
present work:
\begin{itemize}
	\item $^{30.9}$Ga $=>$ Ni or Mn at Ga site,
	\item $^{25.1}$Mn $=>$ Ni or Ga at Mn site,
	\item $^{27.9}$Ni $=>$ Mn at Ni site, and
	\item $^{28.1}$Ni $=>$ Ga at Ni site.
\end{itemize}
All elongated structures of off-stoichiometric Ni$_2$MnGa have been fully
optimized within the VCA approximation. As shown in Figure~\ref{fig:vca}, a
small change in the number of electrons at the Ni site has the most drastic
effect on the energy landscape. The largest difference appears between Ga-rich
($^{28.1}$Ni) and Ni-poor ($^{27.9}$Ni) compounds. Increasing Ni at Ga and Mn
sites lowers the energy minimum at $c/a>1$ compared to the stoichiometric
compound. Conversely, increasing Ga at the Ni site increases the energy of
$c/a>1$ with respect to the stoichiometric compound. These results are in
agreement with previously reported calculations~\cite{APE04}. Differences in the
$E(\delta)$ dependence are more pronounced when the distorted structure is far
from the initial structure. In the next step, the elastic constants of the 
off-stoichiometric Ni$_2$MnGa were calculated using VCA for the tetragonal distorted
structures with $c/a>1$ at the lowest total energy (see Figure~\ref{fig:vca}).

Table~\ref{tab:mech} summarizes the results of the elastic constant calculations
for stoichiometric and off-stoichiometric Ni$_2$MnGa. An extreme effect of the
off-stoichiometry is reflected in the anisotropy ratio ($A_B$). As shown in
Table~\ref{tab:mech}, the $^{27.9}$Ni and $^{30.9}$Ga compounds exhibit negative
anisotropies of -0.02 and -0.4, respectively. In fact, the negative anisotropy
highlights the extreme instability of these systems. In particular, $^{30.9}$Ga
has a large negative value. The results explain why $^{30.9}$Ga (Ga-poor) and
$^{27.9}$Ni (Mn-rich) Ni$_2$MnGa synthesis is difficult. Moreover, the
$^{28.1}$Ni (Ga-rich) and $^{25.1}$Mn (Mn-poor) compounds have large
anisotropies, which are twice the value of the stoichiometric anisotropies. The
large value also explains the tendency of phase transitions in these materials.
Therefore, off-stoichiometric Ni$_2$MnGa --~nearly all synthesized samples are
slightly off-stoichiometric~-- is expected to have pronounced phase transitions
depending on the composition~\cite{CG03}. Thus, deficiency of valence electrons
at the Mn or Ga sites in these systems leads to a negative anisotropy ratio and
thus structural instability. Among the elastic constants, $c_{11}$ and $c_{22}$
are more strongly influenced when the composition changes compared to $c_{66}$
and $c_{44}$, which remain nearly constant. Therefore, a small excess of each of
the elements (small change of the valence electron concentration in the vicinity
of that site) will not change the shear in the [100] direction. However, the
asymmetries $A_{B}$ and $A_{100}$ significantly change compared to $A_{001}$,
reflecting a change of the in-plane chemical bonding. Here, Ni$_{2}$MnGa is used
as an example of the sensitivity of Heusler compounds on their stoichiometry.
Disorder-induced phase transitions have been reported for other Heusler
compounds such as iron-based compounds~\cite{GKK13}.

\subsection{Derived properties of tetragonal Heusler compounds}
\label{prop}
 
Finally, some physical properties and material parameters of the compounds are
derived from the calculated elastic properties. The velocity of sound is an
important quantity. Its averaged values can be directly determined from the
calculated elastic constants. In experiments, on the other hand, the sound
velocities can be used to measure elastic constants. Therefore, the sound
velocities $v$ are nearly synonymous with the elastic stiffness constants $c$.
Further, sound velocities have been used to study various solid-state properties
and processes~\cite{Led06}. Therefore, having the sound
velocities predicted by calculations in advance could be quite important for
experimental measurements. Usually, the directionally dependent acoustic
properties are analyzed in terms of the slowness that is the inverse of the
phase velocity. The group velocities are found from the derivatives of the
slowness.

\begin{figure*}[htb]
\centering
\includegraphics[width=15cm]{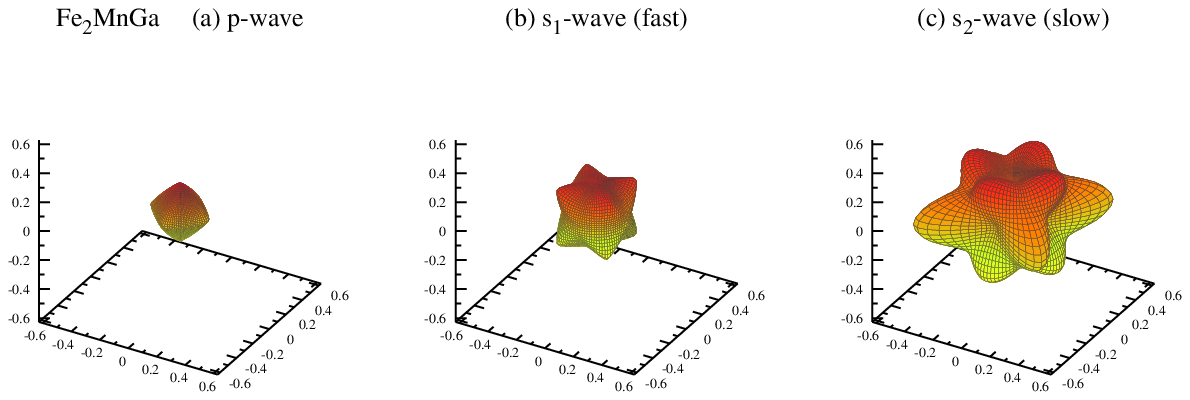}
\includegraphics[width=15cm]{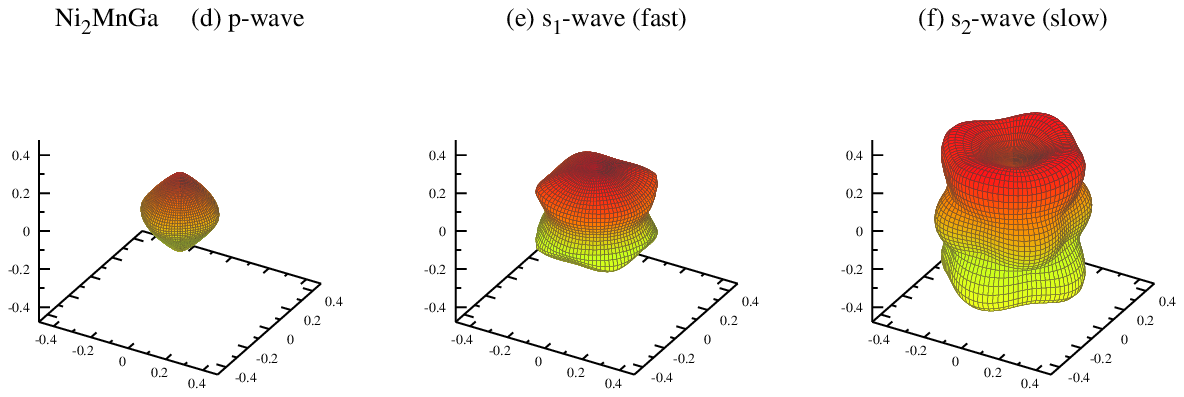}
 \caption{(Online color) Slowness surfaces of Fe$_2$MnGa (a)--(c) and Ni$_2$MnGa (d)--(f). \\
          The slowness is given in (km/s)$^{-1}$. }
\label{fig:slowness_3d}
\end{figure*}

Figure~\ref{fig:slowness_3d} compares the slowness surfaces of Fe$_2$MnGa and
Ni$_2$MnGa. The slowness surfaces reflect the elastic anisotropy in comparison
to Figure~\ref{fig:bmym}, which shows the distribution of Young's modulus. Three
slowness surfaces appear in both cases, representing different polarizations of
the sound wave. The pressure ($p$) wave is longitudinal polarized. Moreover, $p$
has the highest phase velocity and thus the smallest slowness
(Figure~\ref{fig:slowness_3d}(a) and (d)). The remaining two surfaces belong to
the fast ($s_1$) and slow ($s_2$) shear waves that are transversely polarized.
The slowness surfaces of the $p$ waves have a similar shape for both materials,
and their maxima are found along the \{001\}-type principle axes. The shapes of
the slowness surfaces of the shear waves differ between the two compounds. The
observed differences reflect the differences in the anisotropy of both
materials, and it is clearly seen that Ni$_2$MnGa has a much lower anisotropy in
the $x-y$ plane.

Further material parameters are derived from the average sound velocities as
described in Appendix~\ref{app:properties}. At low temperatures, where only
acoustic vibrational modes contribute, the Debye temperature $\theta_{\rm D}$
can be estimated from the average sound velocity~\cite{And63}. The values
estimated in this way are generally larger than Debye temperatures determined
from phonon calculations or in experiments~\cite{QFF012} because the optical
phonon branches are neglected when the elastic constants are used for
calculating the {\it ``acoustical"} Debye temperature 
$\theta_{\rm D}^{\rm acc}$. 
In a similar way, the average sound velocities can be used to estimate the 
{\it ``acoustical"} Gr{\"u}neisen parameter $\zeta^{\rm ac}$ ~\cite{Bel04}.

The calculated average sound velocities together with the estimated Debye
temperatures and Gr{\"u}neisen parameters are listed in Table~\ref{tab:vel}. The
average sound velocities are rather similar for all compounds, ranging from
about 3100 to 3800~m/s. The acoustical Debye temperatures are all above room
temperature, ranging from 376 to 490~K. As expected, compounds consisting of
heavier elements tend to have lower values. With the exception of Mn$_2$NiGa,
all $\zeta^{\rm ac}$ values are about 2. This demonstrates that the
anharmonicity of the lattice vibrations is nearly the same for all compounds.

As further shown in Table~\ref{tab:vel}, the theory and experimental values have
about a 20\% discrepancy for the case of Ni$_2$MnGa. As reported in
Reference~[\onlinecite{CFB08}], however, the stoichiometry of the
compound has a large effect on the measured Debye temperature. As shown in
Table~\ref{tab:mech}, the Debye temperature decreases by about 20\% in
$^{27.9}$Ni. However, in the calculations, the changes in stoichiometry are
extremely small. The changes in the calculated values may be more evident with
larger variations in the stoichiometry. For example, in experiments, the results
change from 261~K in the case of Ni$_{49.6}$Mn$_{21.9}$Ga$_{28.5}$ to 345~K in
the case of Ni$_{53.1}$Mn$_{26.6}$Ga$_{20.3}$. Therefore, having an ideal 2:1:1
system, like that assumed in most theories, is not easily possible or may even
be impossible from an experimental standpoint. However, the estimation of
measurable properties should provide information about the studied system and
its potential for applications.
 
\begin{table}[htb]
\centering
\caption{Derived physical properties of tetragonal Heusler compounds. \newline
         Tabulated are the longitudinal $v_l$, transverse $v_t$, 
         and average $\overline{v}$ sound velocities as well as the {\it acoustical}
         Debye temperature $\theta_{\rm D}^{\rm ac}$ and {\it acoustical} Gr{\"u}hneisen parameter 
         $\zeta^{\rm ac}$ estimated from the sound velocities. 
         All $v$ are given in ms$^{-1}$, $\theta$ is given in K, $\zeta$ is dimensionless.}

  \begin{ruledtabular}
  \begin{tabular}{lccccc}
    Compound  & $v_l$ & $v_t$ & $\overline{v}$ & $\theta_{\rm D}^{\rm ac}$ & $\zeta^{\rm ac}$ \\
  \hline
   Mn$_2$NiGa                        & 5667  & 3441  & 3803 & 490 & 1.67 \\
   Ni$_2$MnGa                        & 5688  & 3219  & 3487 & 451 & 1.95 \\    
   {\it Exp.}\footnotemark[1]        &       &       &      & 345 &      \\
   {\it Other calc.}\footnotemark[2] & 5572  & 2853  & 3196 & 323 &      \\
   Mn$_2$FeGa                        & 5443  & 3023  & 3367 & 434 & 2.02 \\
   Fe$_2$MnGa                        & 5753  & 3253  & 3618 & 470 & 1.96 \\
   Rh$_2$CrSn                        & 5404  & 3078  & 3421 & 410 & 1.93 \\
   Rh$_2$FeSn                        & 5277  & 2803  & 3133 & 376 & 2.19 \\
   Rh$_2$CoSn                        & 5346  & 2885  & 3221 & 389 & 2.13 \\
  \end{tabular}
  \end{ruledtabular}
\footnotetext[1]{Reference~\onlinecite{CFB08}}
\footnotetext[2]{Reference~\onlinecite{OCc10}}
  \label{tab:vel}
  \end{table}

\section{Summary}

In the present work, the elastic constants of tetragonally distorted Heusler
compounds were determined. The full-potential LAPW method together with the
gradient-corrected PBE exchange-correlation functional were employed for all
calculations. The relation between the calculated elastic constants and
convergence criteria were discussed. Increasing only one of the parameters, such
as the $k$-points or $R_{\rm MT}K_{\rm max}$, while keeping the other parameter
low led to large errors in the calculated elastic constants. Therefore, to
calculate both elastic constants accurately, $R_{\rm MT}K_{\rm max}$ and 
$k$-points must be sufficiently large to guarantee convergance. Structural
optimization was shown to have an important effect on the elastic constants for
tetragonal Heusler compounds. The method was used to investigate the crystalline
stability of materials based on the calculation of their elastic properties.

Based on the calculated results, the considered tetragonal Heusler compounds are
intermediate materials, between brittle and ductile. Elastically, they exhibit
mainly metallic rather than covalent bonding. The structural instability,
mechanical properties, structural anisotropy, and other mechanical properties
were also explored. Using the virtual crystal approximation, the importance of
the stoichiometric composition for Ni$_2$MnGa was demonstrated, and extreme
sensitivity on the variation of the Ni component in Ni$_2$MnGa was observed.
Negative anisotropy of $^{27.9}$Ni$_2$MnGa and Ni$_2$Mn$^{30.9}$Ga together with
the large anisotropy of the $^{28.1}$Ni$_2$MnGa and Ni$_2$$^{25.1}$MnGa
compounds indicated instability of off-stoichiometric Ni$_2$MnGa in the
tetragonal phase.

The calculated material properties are useful for applications focusing on bulk
materials. However, the appearance and prediction of metastable tetragonal
structures allows lattice parameter engineering with artificial $c/a$ ratios
initialized by epitaxial or pseudomorphic thin film growth. Thus, Heusler thin
films could be designed to have specific properties.




\appendix
\label{app}

\section{Basic equations for the elastic constants, moduli,
         and related parameters}
\label{app:basic}

The equations that describe the elastic properties of solids have been described
in detail by Nye~\cite{Nye85}; this discussion is summarized here and compares
tetragonal, hexagonal, and cubic systems with focus on the tetragonal case. The strain
matrix $\epsilon$ transforms lattice $A$ with basis vectors $X,Y,Z$ into the
deformed lattice

\begin{equation}
  A'=(\underline{1}+\epsilon)A
\end{equation}

with basis vectors $X',Y',Z'$. The symmetric strain matrix $\epsilon$ contains six 
different strains $e_i$ and has the form:

\begin{equation}
  \underline{1}+\epsilon = \left(
  \begin{array}{ccc} 
        1 + e_1         & \frac{1}{2} e_6 & \frac{1}{2} e_5 \\
        \frac{1}{2} e_6 & 1 + e_2         & \frac{1}{2} e_4 \\
        \frac{1}{2} e_5 & \frac{1}{2} e_4 & 1 + e_3 \\
  \end{array}
  \right)
\label{eq:strain}
\end{equation}

The elastic relations (Hooke's law) between the strain ($\epsilon$) and stress 
($\sigma$) matrices are mediated by the elastic compliance ($\bm S$) or the elastic 
stiffness ($\bm C$) matrices:

\begin{equation}
  \epsilon = {\bm S} \sigma \:\: {\rm or} \:\: \sigma = {\bm C} \epsilon
\label{eq:hook} 
\end{equation}

From the elastic equations, the relations between the compliance matrix and the 
stiffness matrix are

\begin{equation}
  {\bm S} = {\bm C}^{-1}
\label{eq:ctos}
\end{equation}

and vice versa ${\bm C} = {\bm S}^{-1}$. These relations imply that ${\bm {S C}} 
= {\bm {C S}} = \underline{1}$.

In the most general case, the elastic matrix is symmetric and of order $6 \times 6$. 
In triclinic lattices, the elastic matrix contains 21 independent elastic
constants. This number is largely reduced in high symmetry lattices. For
example, in an isotropic system, it contains only the two constants
$c_{11}=c_{22}=c_{33}$ and $c_{12}=c_{13}=c_{23}$, and the remaining diagonal
elements of the matrix are determined by 
$c_{44}=c_{55}=c_{66}=(c_{11}-c_{12})/2$.

In cubic lattices, the three elastic constants $c_{11}$, $c_{12}$, and $c_{44}$
are independent. There are five independent elastic constants for hexagonal
structures ($c_{11}$, $c_{12}$, $c_{13}$, $c_{33}$, and $c_{44}$), while
tetragonal structures have either seven (classes: $4$, $\overline{4}$, or $4/m$)
or six ($c_{11}$, $c_{12}$, $c_{13}$, ($c_{16}=-c_{15}$), $c_{33}$, $c_{44}$ and
$c_{66}$) elastic constants. The elastic matrix for all classes of cubic and
hexagonal crystals as well as the classes $4mm$, $\overline{4}2m$, $422$, or
$4/mmm$ of tetragonal crystals have the form

\begin{equation}
  C = \left(
  \begin{array}{ccc ccc} 
        c_{11}   & c_{12}    & c_{13} &  .      & .      & (c_{16})  \\
        c_{12}   & c_{11}    & c_{13} &  .      & .      & (-c_{16}) \\
        c_{13}   & c_{13}    & c_{33} &  .      & .      & .         \\
        .        & .         & .      &  c_{44} & .      & .         \\
        .        & .         & .      &  .      & c_{44} & .         \\
        (c_{16}) & (-c_{16}) & .      &  .      & .      & c_{66}    \\                                      
  \end{array}
  \right)
\end{equation}

where zero elements are assigned by dots and additional tetragonal elements for
classes $4$, $\overline{4}$, or $4/m$ are given in brackets.
Moreover, the elastic matrix has restrictions $c_{33}=c_{11}$, $c_{66}=c_{44}$, $c_{13}=c_{12}$ 
in cubic systems and $c_{66}=(c_{11}-c_{12})/2$ in hexagonal systems.

The matrix $C^{\rm tetra}$ has six eigenvalues for the classes $4mm$,
$\overline{4}2m$, $422$, or $4/mmm$:
\begin{itemize}
	\item $C_1^t = c_{11}-c_{12}$,
	\item $C_{2,3}^t =\frac{1}{2}\left(c_{11}+c_{33}+c_{12} \pm \sqrt{Z} \right)$ 
	\item $C_4^t =c_{66}$, and 
	\item $C_{5,6}^t =c_{44}$,
\end{itemize}
where $Z = c_{11}^2+2c_{11}c_{12}-2c_{11}c_{33}+c_{12}^2-2c_{12}c_{33}+8c_{13}^2+c_{33}^2$.
The last eigenvalue ($C_{5,6}^t$) is twofold degenerate (also note the double
sign ($\pm$) in the second line). The crystal becomes unstable when one of the
eigenvalues becomes zero or negative or in case that $Z<0$.

The relations between the elastic constants $c_{ij}$ and the elements of the 
compliance matrix $s_{ij}$ are found from Equation~\ref{eq:ctos}. In all classes 
of hexagonal systems or in tetragonal systems belonging to the classes $4mm$, 
$\overline{4}2m$, $422$, or $4/mmm$, the relations between $c_{ij}$ and $s_{ij}$ 
are given by 

\begin{eqnarray}
  s_{11} & = & \frac{c_{11}c_{33} - c_{13}^2}{c (c_{11} - c_{12})}  \\ \nonumber
  s_{12} & = & \frac{-c_{12}c_{33} + c_{13}^2}{c (c_{11} - c_{12})} \\ \nonumber
  s_{13} & = & \frac{-c_{13}}{c}                                    \\ \nonumber
  s_{33} & = & \frac{c_{11} + c_{12}}{c}                            \\ \nonumber
       c & = & c_{33} (c_{11} + c_{12}) - 2 c_{13}^2                \\ \nonumber
  s_{44} & = & \frac{1}{c_{44}}                                     \\ \nonumber
  s_{66} & = & \frac{1}{c_{66}}
\end{eqnarray}

where $c_{66}$ appears only in tetragonal systems. Indeed, the number of
equations is much less in cubic systems as shown from the restrictions given
above.

The elastic properties of single crystals are completely determined by the
elastic matrices ${\bm C}$ and ${\bm S}$. In reality, polycrystalline materials
are considered more often than single crystals. Polycrystalline materials
consist of randomly oriented crystals, and thus, a description of their elastic
properties requires only two independent elastic moduli: the bulk modulus ($B$)
and the shear modulus ($G$). The relationships between the single-crystal
elastic constants and the polycrystalline elastic moduli are given by the
Voigt~\cite{Voi28} or Reu{\ss}~\cite{Reu29} averages. Voigt's approach uses the
elastic stiffnesses $c_{ij}$, while Reu{\ss}'s approach uses the compliances
$s_{ij}$. Voigt's moduli~\cite{Voi28} are given as function of the elastic
constants by the equations:

\begin{eqnarray}
  B_V & = & \frac{1}{9}(2 c_{11} + 2c_{12} + 4 c_{13} + c_{33}) \\ \nonumber
  G_V & = & \frac{1}{15}(2c_{11} - c_{12} - 4 c_{13} + c_{33} + 6 c_{44} + 3 c_{66})
\end{eqnarray}

and Reu{\ss}'s moduli~\cite{Reu29} are usually calculated from the elements of 
the compliance matrix:

\begin{eqnarray}
  B_R & = & \frac{1}{2s_{11} + 2s_{12} + 4s_{13} + s_{33}} \\ \nonumber
      & = & \frac{(c_{11}+c_{12})c_{33} - 2c_{13}^2}{c_{11} + c_{12} + 2 c_{33} - 4 c_{13}} \\ \nonumber
  G_R & = & \frac{15}{8 s_{11} - 4 s_{12} - 8 s_{13} + 4 s_{33} + 6 s_{44} + 3 s_{66}}.
\end{eqnarray}
 
For cubic or isotropic crystals, the bulk moduli in Voigt's ($B_V$) and 
Reu{\ss}'s ($B_R$) approach are equal, as shown by using the 
restrictions on $c_{ij}$ given above. In cases other than isotropic or 
cubic, $G_R$ cannot be easily rewritten in terms of the elastic constants. 

Finally, the mechanical properties of polycrystalline materials are approximated 
in the Voigt--Reu{\ss}--Hill~\cite{Hil52} approach, where the bulk and shear moduli 
are given by arithmetic averages:

\begin{eqnarray}
  B & = & \frac{1}{2}(B_V + B_R) \\ \nonumber
  G & = & \frac{1}{2}(G_V + G_R).
\end{eqnarray}

The bulk modulus $B$ of a material characterizes its resistance to fracture,
whereas the shear modulus $G$ characterizes its resistance to plastic
deformations. Therefore, ratios between the elastic moduli $B$ and $G$ are often
given for characterization and comparison of different materials. Pugh's modulus
$k$ is the simple ratio of the bulk and shear moduli~\cite{Pug54}:

\begin{equation}
   k=B/G.
\label{eq:pugh}
\end{equation}

Poisson's ratio $\nu$ also relates the bulk and shear moduli:

\begin{equation}
   \nu = \frac{1}{2} \ \frac{3B - 2G}{3B + G} = \frac{3k - 2}{6k + 2}.
\label{eq:posison}
\end{equation}

Further, Poisson's ratio bridges between the rigidity modulus $G$ and Young's 
modulus $E$, which is given by

\begin{equation}
  E = 2 G (1+\nu) = \frac{9BG}{3B + G}.
\end{equation}

\section{Elastic stability and representation of elastic properties}
\label{app:stability}

The set of elastic moduli and their ratios allows characterization of the
elastic behavior of materials. However, the mechanical stability is still open.
As a first criterion, the elastic moduli all must be positive. Born and
coworkers developed a theory on the stability of crystal
lattices~\cite{Bor40,*Mis40,*BFu40,*BMi40,*Fue41a,*Fue41b,BHu56}. For tetragonal
crystals at ambient conditions, the seven elastic stability criteria are given
by

\begin{itemize}
	\item $2c_{11}+ c_{33} + 2c_{12} + 4 c_{13} > 0$
	\item $c_{11}+ c_{33} - 2c_{13} > 0$   
	\item $c_{12}, c_{33}, c_{44}, c_{66} > 0$
	\item $c_{11} - c_{12} > 0$
\label{eq:bornhuang}
\end{itemize}

Note that the number of
criteria is reduced for the lower number of elastic constants in hexagonal or
cubic crystals to 5 or 3, respectively. The last condition is used to define the
tetragonal shear modulus $c'=(c_{11}-c_{12})/2$. In some works, the direct
difference $C'=c_{11}-c_{12}$ is used. If external, hydrostatic pressure $p$ is
applied, then the crystal becomes unstable when $2p>C'$, that is, at $p>c'$.

The linear compressibility $\beta$ is the crystal response to hydrostatic
pressure by a length decrease. For cubic systems, the linear compressibility is
isotropic, that is, a sphere of a cubic crystal under hydrostatic pressure
remains a sphere. The situation is different in non-cubic systems where
$\beta=\beta(\hat{r})$ becomes directionally dependent. In hexagonal, trigonal,
and tetragonal systems, the directional dependence is given by

\begin{equation}
  \beta(\hat{r}) = (s_{11}+s_{12}+s_{13}) (\hat{x}^2+\hat{y}^2) - (s_{11}+s_{12}-s_{13}-s_{33})\hat{z}^2.
\end{equation}

The linear compressibility of a cubic crystal is simply $\beta^{\rm cub}=s_{11}+2s_{12}$. 
The volume compressibility $\kappa$ of hexagonal and 
tetragonal systems is also directionally dependent and given in Reu{\ss}'s 
approach by

\begin{eqnarray}
  \kappa(\hat{r}) = &   & (s_{11}+s_{12}+s_{13}) (\hat{x}^2 + \hat{y}^2) \nonumber \\
                    & + & (s_{33}+2 s_{13})\hat{z}^2
\label{kappa}
\end{eqnarray}

For cubic systems, $s_{13}=s_{12}$ and $s_{33}=s_{11}$, that is, 
$\kappa^{\rm cub}= 3(s_{11}+2s_{12})$, and thus, the bulk modulus $B=1/\kappa$ is 
isotropic for crystals with cubic symmetry. For hexagonal and tetragonal 
systems, $\kappa$ becomes isotropic when the two terms $s_{11}+s_{12}+s_{13}$ 
and $s_{33}+2 s_{13}$ in Equation~\ref{kappa} are equal. Therefore, the 
anisotropy of the hexagonal and tetragonal bulk moduli is defined by

\begin{equation}
  A_B = \frac{s_{33}+2 s_{13}}{s_{11}+s_{12}+s_{13}},
\label{eq:anisB}
\end{equation}

and their isotropic compressibility becomes 
$\kappa^{\rm hex, tet}_{\rm iso}=2s_{11}+s_{33}+2s_{12}+4s_{13}$. 

Other than the bulk modulus of cubic crystals, Young's modulus of cubic,
hexagonal, or tetragonal systems is not isotropic. The representation surface of
Young's modulus for tetragonal systems with classes $4mm$, $422$,
$\overline{4}2m$, and $4/mmm$ is given by

\begin{eqnarray}
  \frac{1}{E^t(\hat{r})} = &   & (\hat{x}^4+\hat{y}^4) s_{11} + \hat{z}^4 s_{33} \nonumber \\
                           & + &  \hat{x}^2\hat{y}^2 (2s_{12}+s_{66})            \nonumber \\
                           & + &  \hat{z}^2(1-\hat{z}^2)(2s_{13}+s_{44}).
\end{eqnarray}

For the tetragonal classes $4$, $4/m$, and $\overline{4}$, an additional 
term is present such that

\begin{equation}
  \frac{1}{E^{t'}(\hat{r})} = \frac{1}{E^t(\hat{r})}
                           + 2\hat{x}\hat{y}(\hat{x}^2-\hat{y}^2) s_{16}.
\end{equation}

The shear anisotropic factors provide a measure of the degree of anisotropy in
the bonding between atoms in different planes. The number of different shear
anisotropies depends on the crystal system. In both -- hexagonal and tetragonal
-- systems, the shear anisotropic factors $A_{100}$ (or equivalent $A_{010}$)
for the $\{100\}$ shear planes between the $\left\langle011\right\rangle$ and
$\left\langle010\right\rangle$ directions and $A_{001}$ for the $\{001\}$ planes
between $\left\langle110\right\rangle$ and $\left\langle010\right\rangle$ are:

\begin{eqnarray}
   A_{100} & = & \frac{4 c_{44}}{c_{11} + c_{33} - 2c_{13}} \nonumber \\
   A_{001} & = & \frac{2 c_{66}}{c_{11} - c_{12}}
\label{eq:anis}
\end{eqnarray}

In cubic crystals, both factors are the same 
$A_e=A_{001}= 2 c_{44} / (c_{11}-c_{12})$, 
as mentioned above. In hexagonal systems, $c_{66}=(c_{11} - c_{12})/2$,
and thus, $A_{001}=1$. For isotropic crystals, all $A$ factors must be unity,
while any value smaller or greater than unity is a measure of the degree of
elastic anisotropy possessed by the crystal.

Comparing the equations~(\ref{eq:anisB},\ref{eq:anis}) for the elastic
anisotropies with the Born--Huang~\cite{BHu56} criteria, these equations can
clearly be used to show the elastic stability. Most obviously, crystals with one
negative anisotropy are not stable. Further, crystals with large anisotropies
also tend to instabilities; in particular, crystals are not stable for
$A\rightarrow\infty$ when one of the denominators becomes zero. This behavior
makes the anisotropies important parameters, even though they may not cover all
possible causes for Born--Huang instabilities.

\section{Equations for calculating properties from the elastic constants}
\label{app:properties}

Besides the elastic moduli, further important physical quantities can be derived
from the elastic constants. Acoustical spectroscopy is widely used to determine
the elastic properties of crystalline solids. The propagation of sound waves in
solids is described by the Christoffel equation:

\begin{equation}
	(\Gamma_{ij} - \rho v^2 \delta_{ij}) U_j = 0,
\end{equation}

where $v$ is the phase velocity, $\rho$ is the mass density, $\delta_{ij}$ is the
Kronecker delta, $U$ is the polarisation vector, and

\begin{equation}
     \Gamma_{ij}=c_{ijkl}l_jl_l
\end{equation}

is the Christoffel tensor built from the elastic constants and the direction
cosines $l_i$ ($i=1\ldots3$) that describe the direction of wave motion.
For tetragonal systems, the Christoffel tensor is given by

\begin{eqnarray}
 \Gamma_{11} & = & c_{11} l_1^2 + c_{66} l_2^2 + c_{44} l_3^2 + 2 c_{16} l_1l_2   \\  \nonumber
 \Gamma_{22} & = & c_{66} l_1^2 + c_{11} l_2^2 + c_{44} l_3^2 - 2 c_{16} l_1l_2   \\  \nonumber
 \Gamma_{33} & = & c_{44} l_1^2 + c_{44} l_2^2 + c_{33} l_3^2   \\  \nonumber
 \Gamma_{12} & = & c_{16} l_1^2 - c_{16} l_2^2 + (c_{12} + c_{66}) l_1l_2 - c_{16} l_2l_3   \\  \nonumber
 \Gamma_{13} & = & (c_{13} + c_{44}) l_1l_3   \\  \nonumber
 \Gamma_{23} & = & (c_{13} + c_{44}) l_2l_3
\end{eqnarray}

and $\Gamma_{ij}=\Gamma_{ji}$. The Christoffel tensor reduces for the classes $4mm$, 
$\overline{4}2m$, $422$, and $4/mmm$, where $c_{16}=0$, to

\begin{eqnarray}
 \Gamma_{11} & = & c_{11} l_1^2 + c_{66} l_2^2 + c_{44} l_3^2   \\  \nonumber
 \Gamma_{22} & = & c_{66} l_1^2 + c_{11} l_2^2 + c_{44} l_3^2   \\  \nonumber
 \Gamma_{33} & = & c_{44} l_1^2 + c_{44} l_2^2 + c_{33} l_3^2   \\  \nonumber
 \Gamma_{12} & = & (c_{12} + c_{66}) l_1l_2   \\  \nonumber
 \Gamma_{13} & = & (c_{13} + c_{44}) l_1l_3   \\  \nonumber
 \Gamma_{23} & = & (c_{13} + c_{44}) l_2l_3
\end{eqnarray}

The solution of the characteristic $3\times3$ matrix results in a third-order
equation in $v^2$ for the phase velocity. Three distinct modes appear, one with
longitudinal and two with transversal polarisation. Due to possible mixing,
these modes are often referred to as {\it quasi}-longitudinal or 
{\it quasi}-transversal modes. The longitudinal mode corresponds to a pressure ($p$-wave) or
compression wave as it appears also in gases. On the other hand, the transversal
modes appear for solids, and they are distinguished as fast ($s_1$) and slow
($s_2$-wave) shear waves. The wave properties are presented as slowness
surfaces.

The elastic constants also allow direct estimation of the averaged sound 
velocity $\overline{v}$ from the longitudinal ($v_l$) and transverse ($v_t$) 
elastic wave velocities of isotropic materials, which are given by

\begin{eqnarray}
  v_l & =  & \sqrt{ \frac{3B+4G}{3\rho} }   \\ \nonumber
  v_t & =  & \sqrt{ \frac{G}{\rho} },
\end{eqnarray}

where $\rho$ is the mass density of the material. Here, $\overline{v}$ is 
approximately predicted by 

\begin{equation}
  \overline{v} = \left[ \frac{3}{v_l^{-3} + 2 v_t^{-3}} \right] ^{1/3}.
\end{equation}

For low temperatures, where only acoustic vibrational modes contribute, the
Debye temperature $\Theta_D$ can be estimated from the average sound velocity
using the relation~\cite{And63}:

\begin{equation}
  \Theta^{\rm ac}_D = \overline{v} \frac{h}{k_B} \sqrt[3]{ \frac{f}{4\pi} \frac{N_A \rho}{M} }
                    = \overline{v} \frac{h}{k_B} \sqrt[3]{ \frac{f}{4\pi} \frac{1}{V_p} } ,
\end{equation}

where $h$, $k_B$, and $N_A$ are Plank's constant, Boltzman's constant, and Avogadoro's 
number, respectively. The degree of freedom for $n$ atoms in a 
primitive cell with volume $V_p$ ($f=12$ for Heusler compounds with $L2_1$ 
structure) is $f=3n$, and $M$ is the molecular mass, that is, the sum of
all masses of the atoms in the primitive cell of the compound.

In solids, the Gr{\"u}neisen parameter $\zeta$ is a measure of the anharmonicity
of the interactions between the atoms. In general, it is calculated from
logarithmic derivatives of the vibrational frequencies with respect to the
crystal volume. However, full phonon calculations as function of crystal volumes
are demanding tasks, and fast estimates are thus welcome.
Belomestnykh~\cite{Bel04} derived an {\it ``acoustical"} Gr{\"u}neisen
parameter $\zeta^{\rm ac}$ that is directly related to the sound velocities.
Therefore, $\zeta^{\rm ac}$ is given by

\begin{equation}
  \zeta^{\rm ac} = \frac{3}{2} \frac{(3v_l^2 - 4 v_t^2)}{(v_l^2 + 2v_t^2)}.
\end{equation}

\bibliography{elastic_tetra}

\end{document}